\documentclass[pra,twocolumn,showpacs,letterpaper,superscriptaddress]{revtex4-1}

\usepackage{graphicx,amsmath,amssymb,amsfonts,latexsym,color,dcolumn,bm,epsfig,subfigure,pgf}
\usepackage[plainpages=false,hyperfootnotes=false,colorlinks=false]{hyperref}
\usepackage[normalem]{ulem}
\usepackage{dsfont}

\renewcommand{\imath}[0]{\mathrm{i}}

\begin{document}

\title{Polaritonic Contribution to the Casimir Energy between two Graphene Layers}

\author{C.\ H. Egerland}
\affiliation{Humboldt-Universit\"at zu Berlin, Institut f\"ur Physik,
AG Theoretische Optik \& Photonik, 12489 Berlin, Germany}
\affiliation{Max-Born-Institut, 12489 Berlin, Germany}

\author{K. Busch}
\affiliation{Humboldt-Universit\"at zu Berlin, Institut f\"ur Physik,
AG Theoretische Optik \& Photonik, 12489 Berlin, Germany}
\affiliation{Max-Born-Institut, 12489 Berlin, Germany}

\author{F. Intravaia}
\affiliation{Humboldt-Universit\"at zu Berlin, Institut f\"ur Physik,
AG Theoretische Optik \& Photonik, 12489 Berlin, Germany}

\begin{abstract}
We study the role of surface polaritons in the zero-temperature Casimir effect
between two graphene layers that are described by the Dirac model. A parametric
approach allows us to accurately calculate the dispersion relations of the relevant
modes and to evaluate their contribution to the total Casimir energy.
The resulting force features a change of sign from attractive to repulsive as
the distance between the layers increases. Contrary to similar calculations
that have been performed for metallic plates, our asymptotic analysis demonstrates
that at small separations the polaritonic contribution becomes negligible relative
to the total energy.
\end{abstract}

\maketitle

\section{Introduction}
\label{sec:introduction}

As technology progresses further towards miniaturization, effects that are usually
imperceptible at large scales start to become important. Prominent examples are
dispersion forces and in particular the Casimir effect which, in its simplest form,
describes an attractive interaction between two electrically neutral nonmagnetic
macroscopic objects placed in vacuum at zero temperature.
In the approach originally followed by Casimir in 1948
\cite{casimir48},
the force was derived by summing the zero-point energies associated with the
electromagnetic modes of the system. In the case of an empty cavity formed by
two parallel material interfaces, the Casimir energy is given by
\begin{equation}
    E = \sum_{\sigma, n} \sum_{\textbf{k}} \left[ \frac{\hbar}{2}\omega_n^{\sigma}(\mathbf{k},L) \right]_{L \rightarrow \infty}^L,
    \label{eq:casSum}
\end{equation}
where $\hbar$ is the reduced Planck constant, $\sigma = \textrm{TE},\textrm{TM}$
indicates the field polarization and $\textbf{k}=(k_x,k_y)$ is the component of
the wavevector parallel to the interfaces.
The bracket notation describes the regularization procedure introduced by Casimir
to extract a finite result and implies the difference between the sum evaluated at
a finite separation $L$ of the material interfaces and the same sum calculated
in the limit $L\to \infty$.
Physically this amounts to setting the zero of the energy to correspond to a
configuration where no interaction between the objects occurs.

While in the summation every mode is treated equally, this does not mean that
their relative contributions to the final result are equal, too. In fact, in
earlier calculations it was pointed out that, for the Casimir effect, surface
polariton modes play a special role
\cite{van-kampen68,henkel04,intravaia05,barton05,bordag06,intravaia07,haakh16}.
These surface polaritons are mixed light-matter excitations that exist at the
interface of two media and are usually associated with electromagnetic fields
which decay exponentially away from the surface
\cite{economou69,joulain05,pitarke07,maier07}.
Over the last decade, these solutions of the Maxwell equations have attracted
considerable interest due to their unique properties and the possibilities they
offer to nano-photonics and (quantum) optical technologies
\cite{nie97,ebbesen98,maier07,torma15}.
Specifically, in the case of two metallic plates it has been shown that the
surface plasmon polaritons dominate the Casimir interaction at short distances
and strongly affect the force at large distances. This suggests to control the
Casimir effect by manipulating the surface polaritons' properties via
structuring the surface
\cite{Intravaia12a,intravaia13}.

Other methods to tailor the interaction usually rely on the optical properties
of the materials comprising the objects. For instance, in recent studies
graphene has emerged as an interesting candidate and, due to its exotic properties,
is being considered both in theoretical and experimental research
\cite{fialkovsky11,banishev13,Biehs14,klimchitskaya15,klimchitskaya15a,bordag16,Henkel18,Klimchitskaya18a}.
In fact, this research is also relevant in connection to the role played by
dispersion forces in the context of so-called van der Waals materials
\cite{Geim13,Gobre13,Novoselov16,Ambrosetti16,Woods16}.
For such investigations, an adequate theoretical description of graphene's
optical properties is important in order to predict the right magnitudes of
the Casimir force at the relevant length scales
\cite{klimchitskaya15}.
One of the most successful corresponding material model is the so-called Dirac
model, which describes the collective motion of the electrons in graphene in
terms of a (2+1) Dirac field
\cite{mccann12}.

In this manuscript, we merge the previous perspectives and analyze the role of
the surface polaritons for the Casimir effect between two graphene layers that
are described by the Dirac model.
In our approach, we consider the case where graphene might feature a small gap
$\Delta$ in its band structure
\cite{klimchitskaya15,Werra16},
due to, for example, strain or impurities
\cite{Zhou07,Giovannetti07,Chen14,Jung15}.
In Sec.~\ref{sec:dirac} we start by analyzing the behavior of the scattering
coefficients for a single graphene layer described within the Dirac model.
We then calculate the total Casimir energy at zero temperature and determine
its behavior for short and large separations between the layers (Sec.~\ref{sec:casEnergy}) .
Based on this analysis, we proceed in Sec.~\ref{sec:polModes} to calculating the dispersion
relation of the polaritonic modes and, in Sec.~\ref{sec:contribution}, their
contribution to the Casimir energy.
Specifically, we contrast the asymptotic behavior of the polaritonic contribution
to that of the total energy in order to highlight the analogies and the differences
with respect to the result obtained for ordinary metals \cite{intravaia07}.
In Section~\ref{sec:conclusion}, we discuss our findings.

\section{Dirac Model for Graphene}
\label{sec:dirac}

The Dirac model describes the electronic excitations in graphene as fermions
moving in (2+1) spacetime dimensions at the Fermi velocity
$v = v_{\mathrm{F}}/c \approx 1/300$, where $c$ is the speed of light in vacuum.
This effective description is valid up to an energy $E_{\rm Max}$ of a few eV.
Therefore, this sets a frequency limit $\omega_{\rm Max}$ of hundreds of terahertz
beyond which the reliability of the results for graphene's optical response obtained
within this approach becomes questionable.
Pristine graphene corresponds to massless fermions
\cite{mccann12}.
However, previous work has shown that graphene's band structure may feature a
band gap $\Delta \approx 5 - 50 \text{meV}$
\cite{Zhou07,Giovannetti07,Chen14,Jung15},
which can be modeled as an effective mass in the (effective) Dirac equation.
This introduces an additional scale into our system which, in terms of a Compton-like wavelength,
is given by $\lambda_{\Delta} = \hbar c/2 \Delta$.
For the values of the gap mentioned above we have that $\lambda_{\Delta} \approx 2 - 20\, \mu$m.

Within this description, the scattering (reflection and transmission)  coefficients
for a single graphene layer can be obtained by solving a spinor loop diagram in
the aforementioned (2+1) dimensions and subsequently coupling the emerging
polarization tensor to the electromagnetic field
\cite{fialkovsky11,chaichian12}.
Following this approach, the reflection coefficients can be written as
\begin{subequations}
    \begin{gather}
        r_{\mathrm{TM}} (\omega, k) = \frac{\kappa \underline{\Pi}_{00}}{\kappa \underline{\Pi}_{00} + 2k^2}, \\
        r_{\mathrm{TE}} (\omega, k) = \frac{k^2 \underline{\Pi}_{\mathrm{tr}} - \kappa^2 \underline{\Pi}_{00} }{k^2 \left(\underline{\Pi}_{\mathrm{tr}} + 2 \kappa \right)-\kappa^2 \underline{\Pi}_{00}},
    \end{gather}
\end{subequations}
where $k = |\mathbf{k}|=\sqrt{k_x^2+k_y^2}$ and $\kappa = \sqrt{k^2 - \omega^2/c^2} = -\mathrm{i} k_z$
is connected to the wavevector component perpendicular to the plane.
The definition of the square root is chosen so that $\mathrm{Im}[\kappa] < 0$
and $\mathrm{Re}[\kappa] \ge 0$: $\kappa$ is real for evanescent waves and
imaginary in the propagating sector.
$\underline{\Pi}$ is the polarization tensor, where $\underline{\Pi}_{00}$
denotes its ``00" entry (the index ``0'' corresponding to the time dimension)
and $\underline{\Pi}_{\mathrm{tr}}$ denotes the polarization tensor's trace
(over the temporal index ``0'' and the spatial indices ``1'' and ``2'') \cite{fialkovsky11}.
For arbitrary temperature, nonzero chemical potential and a nonzero band
gap, the polarization tensor and its components feature rather complicated
expressions
\cite{fialkovsky11,chaichian12}.
The previous equations for the reflection coefficients take into account
the nonlocal interaction between the electromagnetic field and the electrons
in graphene and, in the appropriate limits, reduce to the expressions
in the so-called optical approximation
\cite{Falkovsky07,Falkovsky08}.
For simplicity, we restrict ourselves to the case of zero temperature ($T=0$)
and undoped sheets,
corresponding to a vanishing chemical potential.
It is also convenient to define the dimensionless quantities
$\lambda = L/\lambda_{\Delta}$,
$K = k\lambda_{\Delta}$,
$\Omega = \omega \lambda_{\Delta}/c$
and
$\mu = \kappa\lambda_{\Delta} = \sqrt{K^2 - \Omega^2}$.
Within this notation, the entries of the polarization tensor take on a more
compact form and we can write
\begin{subequations}
    \begin{gather}
        \underline{\Pi}_{00}(\Omega, K, \Delta) = 2 \frac{\alpha}{\lambda_{\Delta}} \frac{K^2}{p^2} \psi \left(p  \right), \\
        \underline{\Pi}_{\mathrm{tr}}(\Omega, K, \Delta) = 2 \frac{\alpha}{\lambda_{\Delta}} \frac{\mu^2 - p^2}{p^2} \psi \left( p \right),
    \end{gather}
\end{subequations}
where $\alpha$ is the fine-structure constant, $p = \sqrt{\Omega^2 - v^2 K^2}$
and $\psi (p) = \left(p + 1/p\right) \text{arctanh}(p) -1$
\cite{bordag15}.
The function $\psi (p)$ is positive for $0<p<1$. For $p\to 0$, it behaves as
$\psi (p)\approx 4p^{2}/3$  and it diverges for $p\to 1$. The value $p =1$ corresponds
to an effective pair-creation threshold and physically corresponds to the case
where the energy of the (2+1) Dirac field equates the gap (i.e. the
effective mass).
In the $\Omega$-$K$-plane the pair-creation threshold regime is represented by
the curve $ \Omega_{\mathrm{pc}}(K) =\sqrt{1+v^2 K^2} $ (see Fig.~\ref{fig:modes}).
For $p > 1$ the function $\psi (p)$ as well as the reflection coefficients become
complex quantities, indicating the conversion of some of the energy in electron-hole
pair excitations. The limit $p\gg 1$ is equivalent to the case $\Delta\to 0$ and
gives $\psi(p)\to \imath p \pi/2$.

For zero temperature and undoped sheets
the reflection coefficients take the form
    \begin{equation}
        r_{\mathrm{TM}} (\Omega, K) = \frac{\alpha \mu \psi(p)}{\alpha \mu \psi(p) + p^2},
        r_{\mathrm{TE}} (\Omega, K) = -\frac{\alpha \psi(p)}{\alpha \psi(p)-\mu}.
 \label{eq:refcoeffs}
    \end{equation}
Here, their dependence on $\Omega$ and $K$ is implicitly captured via the parameters
$\mu$ and $p$.
As expected on physical grounds, the above expressions show that for our system
the scattering of the electromagnetic radiation is controlled in strength via
the fine-structure constant $\alpha$ and, therefore, is generally weaker than
for ordinary materials.
The above reflection coefficients do not fulfill the ultraviolet transparency
condition and do not vanish in the limit $\Omega\to \infty$, where $r\sim \alpha$.
However, as discussed above, it is well-known that in this limit the Dirac model
becomes unreliable, a characteristic feature which has to be taken into account
when interpreting corresponding calculations.

\section{Casimir Energy of Graphene}
\label{sec:casEnergy}

Before we focus on the contribution of the polaritonic modes it is useful to analyze the
behavior of the total Casimir energy per area.
It is given by the Lifshitz formula
\cite{lifshitz56}
which for our system reads
\begin{equation}
    E(L)/A = \hbar \int\limits_{0}^{\infty} \frac{\mathrm{d} \xi}{2\pi} \int \frac{\mathrm{d}^{2}\mathbf{k}}{(2\pi)^{2}}  \sum_{\sigma} \ln \left[1- r^{2}_{\sigma}(i\xi,k) e^{-2 \kappa L}\right],
    \label{eq:lifshitz}
\end{equation}
where $r_{\sigma}(i\xi,k)$ are the reflection coefficients evaluated along the positive
imaginary frequency axis in the complex $\omega$-plane and $A$ denotes the area of the layers.
For our purposes and in analogy to the procedure followed in previous works
\cite{genet03,intravaia07},
it is convenient to introduce the correction factor
\begin{equation}
    \eta = E(L)/ E_{\rm perf}(L),\quad E_{\rm perf}(L)=-\frac{\hbar c\pi^{2}}{720}\frac{A}{L^{3}},
\end{equation}
which describes the impact of the material properties with respect to the expression for
the Casimir energy between two perfectly reflecting surfaces, $E_{\rm perf}(L)$.
Since the perfect electric conductor limit represents an upper bound for the Casimir effect
between two identical material layers,
$\eta \le 1$ indicates that realistic material properties lead to an interaction with reduced
strength.
In general, the correction factor depends on the system's parameters and, at short layer
separations and for ordinary materials, it goes to zero $\propto L$, describing the transition
from the retarded ($\propto L^{-3}$) to the nonretarded (van der Waals) limit of the Casimir
energy ($\propto L^{-2}$)
\cite{intravaia07}.
For ordinary materials and for large values of $L$ the correction factor tends to a constant,
showing that, in the case of real materials, Casimir's result for perfect reflectors is
simply reduced by a prefactor (at large separations $\eta\to 1$ for metals).
In the case of graphene, using dimensionless variables, we have that $\eta\equiv\eta(\alpha,v,\lambda)$.
However and contrary to the case of ordinary materials, in the short distance limit
($\lambda\to 0$) the correction factor tends to a constant given by
\begin{equation}
   \eta(\alpha,v,0)\approx \frac{45}{\pi^{4}}\left[g_{\mathrm{TM}}(\alpha,v) + g_{\mathrm{TE}}(\alpha,v) \right],
\label{eq:eCasShort}
\end{equation}
where $g_{\sigma}(\alpha,v)$ are involved functions whose details are given in Appendix~\ref{sec:casEnergCalc}.
For values of $\alpha \ll 1$ we see that $\eta$ scales as $\propto \alpha^2$, with
a proportionality factor that depends on $v$. Conversely, for small values of the Fermi
velocity $\eta$ tends to a constant that depends on $\alpha$ (see Appendix~\ref{sec:casEnergCalc}).

\begin{figure}
    \resizebox{\linewidth}{!}{\includegraphics{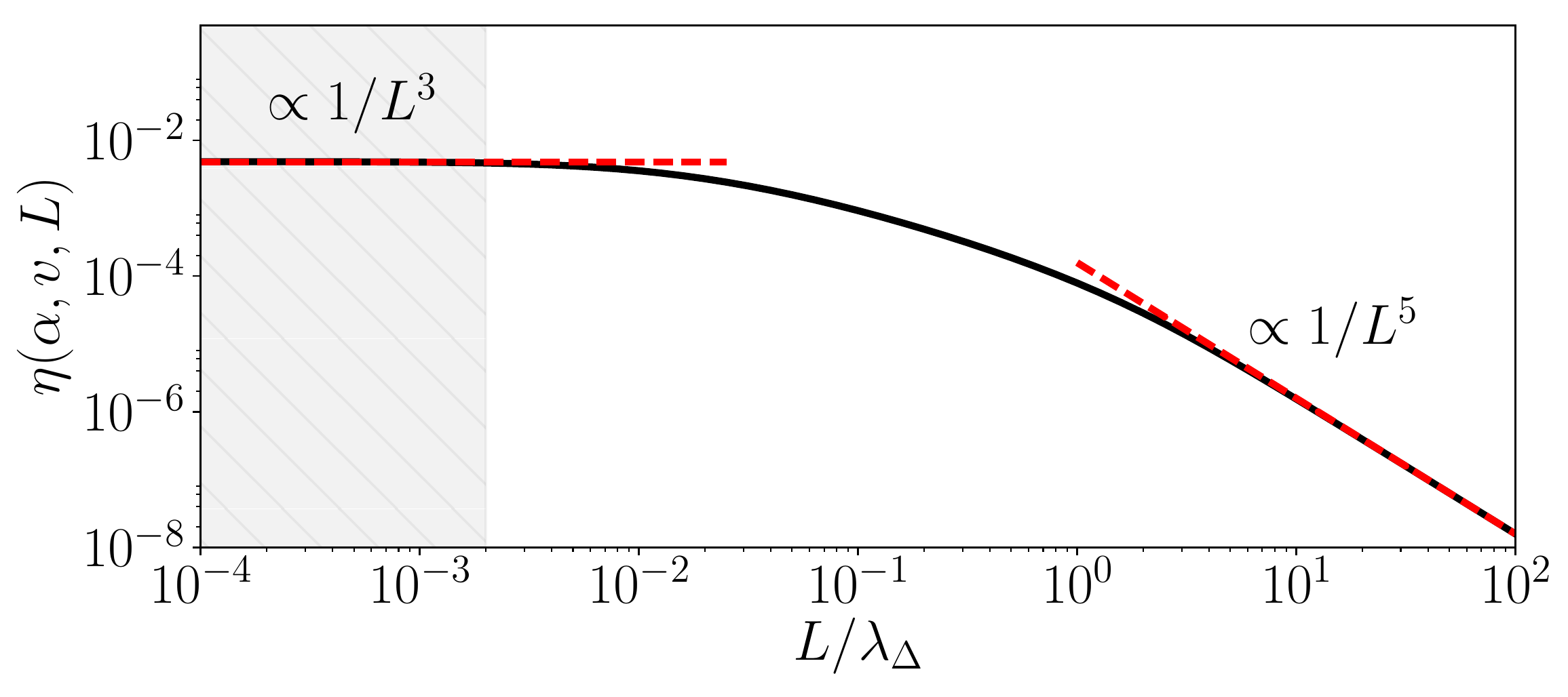}}
    \caption{
        Correction factor $\eta$ as a function of the separation length $L$ measured in units
				of $\lambda_{\Delta}$. We have used the values $\alpha=1/137$ and $v=1/300$.
                The asymptotes, shown as red dashed lines, illustrate a scaling (of the Casimir energy) of $\propto L^{-3}$
				for short separations ($L\ll \lambda_{\Delta}$) and $\propto L^{-5}$ for
				large separations ($L\gg \lambda_{\Delta}$).
        The gray hatched area corresponds to the regime for which the predictions of
				the Dirac model become less reliable -- where we have choses
				$\lambda_{\rm Min}/\lambda_{\Delta} = 2 \times 10^{-3}$.
    }
    \label{fig:eCas}
\end{figure}

The above result has to be considered with care since it is connected with the behavior
of the optical response of graphene in a frequency region where a description in terms
of the Dirac model starts to fail.
The corresponding constraint corresponds to a minimal distance $\lambda_{\rm Min}=c/\omega_{\rm Max}$
below which the above results start to become inaccurate. In Fig.~\ref{fig:eCas}, we
mark this regime with a gray shading.
Still, the value $\lambda_{\rm Min}$ is about two to three orders of magnitude smaller
than $\lambda_{\Delta}$ given above. The expression of $\eta(\alpha,v,0)$ does not depend
on the size of the gap and it is, therefore, equivalent to its value for $\Delta=0$.
Equation \eqref{eq:eCasShort} is, therefore, in agreement with the $\propto L^{-3}$
scaling of the Casimir energy that, in the limit of zero band gap, has previously
been observed for all finite separations
\cite{klimchitskaya13}.
For realistic values of $\alpha$ and $v$ we obtain $\eta(\alpha,v,0) \approx 4.8 \times 10^{-3}$,
indicating a reduction of three orders of magnitude relative to the perfect reflector
case.
We also note that the contribution of the TM mode (quantified by $g_{\mathrm{TM}}$)
accounts for 99.6\% of the value of $\eta(\alpha,v, 0)$, showing the significance of
this polarization to the overall Casimir interaction in the case of graphene.

In the limit of large separations, i.e. for $\lambda \gg 1$, we have instead
\begin{equation}
    \eta(\alpha,v,\lambda) \approx \frac{240 \alpha^2}{\pi^4 \lambda^{2}}\left[ 1 + \frac{1}{15}\left( 3 + 4v^2 + 3v^4 \right) \right],
\end{equation}
indicating a change in power-law behavior of the energy from $\propto L^{-3}$ to
$\propto L^{-5}$ and showing that the presence of a band gap leads to a change of
the Casimir force's scaling that is accompanied by a reduction in magnitudes.
This can be understood by considering that, at large separations, the Casimir
effect effectively probes the low frequency optical response of the material:
The presence of a band gap makes graphene a poor reflector at low energies.
This behavior is, however, very different with respect to that of ordinary metals
(which, for low frequencies, act as nearly perfect reflector) and explains the
deviation from the $\propto L^{-3}$ power law.
Still, the change in the exponent of the power-law is unusual: For ordinary
materials, in going from the nonretarded to the retarded limit, the exponent
usually changes by one unit due to the occurrence of the length scale provided
by the plasma-frequency of the medium.
This variation of two units in the exponent can, once again, directly be attributed to the different behavior
of graphene's reflection coefficients: For a nonzero gap, these coefficients
feature a dielectric-like behavior with a reflectivity that vanishes in the
limit $k,\omega\to0$, while for ordinary metals it tends to a constant
(see Appendix~\ref{sec:casEnergCalc}).

\section{Polaritonic Modes}
\label{sec:polModes}

For a single planar object, surface polariton modes are associated with resonances
in the corresponding reflection coefficients.
Consequently, their dispersion relation can be determined by solving $ r^{-1}_{\sigma} (\omega, k)=0$.
Since $\psi(p)$ is real for $p<1$ and larger than zero for $0<p<1$, we can infer from
the expressions in Eq. \eqref{eq:refcoeffs} that, contrary to the usual behavior of
ordinary metals, polaritonic modes only appear in the TE polarization. This leads to
profound modifications of the electromagnetic field profiles that are connected with
these excitations.
For ordinary metals, the polaritonic resonances are typically associated with a
TM-polarized field, which is predominantly electric.
This property is related to the charge oscillations bounded to the surface (plasmons
for metals) constituting the matter part of the polaritonic mode. Instead, a TE-polarized
field is predominantly magnetic in nature.
This behavior is connected with the statistically induced change in sign of the spinor
loop which gives the polarization tensor for graphene
\cite{bordag14,bordag15}.
In Ref.
\cite{mikhailov07},
where TE resonances were analyzed, this behavior has been associated with a change
in sign of the interband contribution to graphene's conductivity with respect to the
behavior of the intraband conductivity of ordinary metals.

\begin{figure}
    \centering
    \resizebox{\linewidth}{!}{\includegraphics{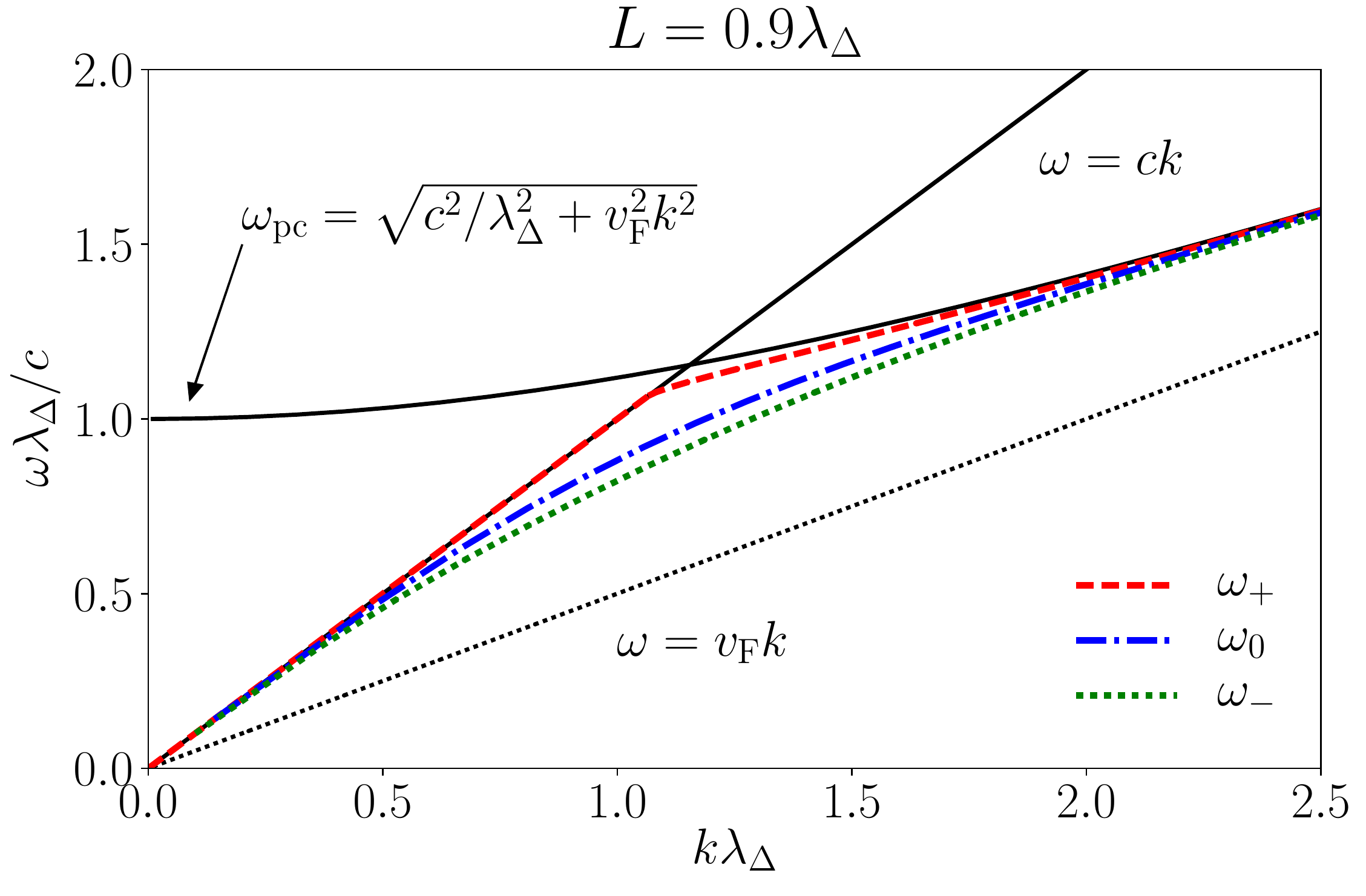}}
    \caption{
        The dispersion relation of the single layer polariton $\omega_0(k)$ (blue dash-dotted line)
				and of the two coupled polaritons $\omega_\pm(k)$ (red dashed and green dotted lines) all lie
				below the pair-creation threshold $\omega_{\rm pc}(k)$.
        For better graphical representation the dispersion relations are plotted for
				the parameters $\alpha=v=1/2$.
        The antisymmetrically coupled polariton $\omega_+(k)$ (red dashed line) has to be
				continued with $\omega=ck$ starting from the point where it becomes
				tangent to the light cone, downward to zero.
 \label{fig:modes}   }
\end{figure}

The dispersion relation for the TE-polarized surface polariton of a single graphene
layer can, in the $\Omega$-$K$ plane, be described in terms of the parametric curve
\begin{equation}
    \Omega_{0}(K)\equiv
    \begin{cases}
        \Omega_0(\mu) =\sqrt{\frac{v^2\mu^2 + \psi^{\rm iv}\left[\frac{\mu}{\alpha}\right]^2}{1-v^2}}\\
        K_0(\mu) = \sqrt{\frac{\mu^2 + \psi^{\rm iv}\left[\frac{\mu}{\alpha}\right]^2}{1-v^2}}
    \end{cases},
    \label{eq:singlemode}
\end{equation}
where $\psi^{\rm iv}$ denotes the inverse function of $\psi$, i.e. $\psi(\psi^{\rm iv}(x))=x$.
Since for $v  < 1$ one has that $\Omega_0(\mu) \le K_0(\mu) $, indicating that the
field associated with this surface polariton is evanescent in agreement with  $\mu \in (0,\infty)$.
Furthermore, the resulting dispersion relation is bounded from above by the pair-creation
threshold frequency $\Omega_{\mathrm{pc}}(K)$.
These features are also clearly visible in Fig.~\ref{fig:modes}.
In particular, we observe that the polaritonic dispersion curve for the single graphene
layer lies entirely below the light cone ($\omega=ck$ or equivalently $\mu=0$), goes
to zero for small wavevectors and tends to the pair-creation frequency for large $k$.
The behavior of the TE-polarized surface polariton has already been examined in great
detail in the existing literature using a semi-analytical approach in Ref.
\cite{bordag14}
and using a parametric representation in Ref.
\cite{Werra16}.

\begin{figure}
    \centering
    \includegraphics[width=0.6\linewidth]{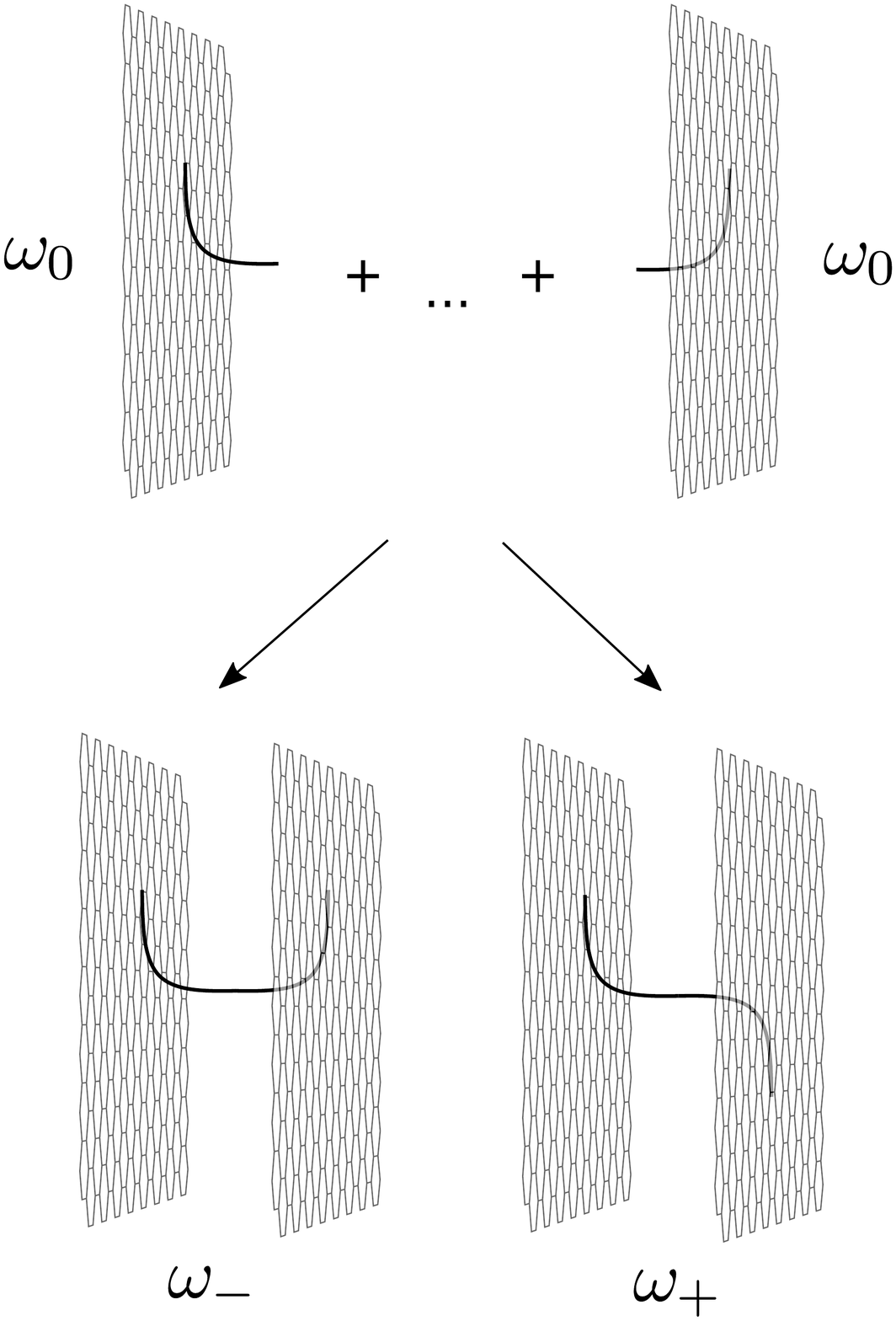}
    \caption{
        Schematic illustration of the setup analyzed in this work.
        The Casimir force between two identical, parallel, undoped graphene layers is considered.
        Particular attention is devoted to the polaritonic contribution.
        When the distance between the layers is reduced, the polaritons featured by each
				of the graphene layers start to interact.
        This interaction removes the degeneracy and the coupled modes are distinguished in
				terms of there associated field into a symmetric ($\omega_{-}$) and antisymmetric
				($\omega_{+}$) coupled surface polariton.
    \label{fig:polariton} }
\end{figure}

In the case of two identical, parallel graphene sheets, the polaritonic excitations that live
on each layer couple through their evanescent tails (see Fig.~\ref{fig:polariton}).
In this case the dispersion relation of the corresponding coupled modes can be found from
the solutions of
\begin{equation}
    1 - r_{\mathrm{TE}}^2(\omega, k) e^{-2 \kappa L} = 0 \Rightarrow -r^{-1}_{\mathrm{TE}}(\omega, k) = \pm e^{-\mu \lambda}.
    \label{eq:resoBi}
\end{equation}
The solutions we are looking for must, in the limit $\lambda\to \infty$, tend to
Eq.~\eqref{eq:singlemode} since in this case the two sheets do not interact and the single
layer case must be recovered.
At finite separations, the interaction removes the degeneracy and two distinct coupled polaritons
arise with a dispersion relation that also depends on the distance between the layers through
the parameter $\lambda$.
These coupled polaritons can be classified in terms of the properties of their electromagnetic
field and we distinguish between an antisymmetric ($+$ sign) and a symmetric ($-$ sign)
polaritonic excitation.
Similar to the single layer case, their dispersion relations are given in terms of the
parametric expressions
    \begin{equation}
    \Omega_{\pm}(K,\lambda)\equiv
      \begin{cases}
        \Omega_{\pm}(\mu, \lambda) = \sqrt{\frac{v^2\mu^2 + \psi^{\rm iv}\left[\frac{\mu}{\alpha}f_{\pm}(\mu \lambda)\right]^2}{1-v^2}}
    \\
        K_{\pm}(\mu, \lambda) =  \sqrt{\frac{\mu^2 + \psi^{\rm iv}\left[\frac{\mu}{\alpha}f_{\pm}(\mu \lambda)\right]^2}{1-v^2}}
        \end{cases},
     \label{eq:bimodes}
    \end{equation}
where we have defined the function $f_{\pm}(x) = (1 \mp e^{-x})^{-1}$.
Since $f_{\pm}(x\to \infty)\sim 1$ at large separation, i.e.\ $\lambda \gg 1$, Eq.~\eqref{eq:bimodes}
approaches Eq.~\eqref{eq:singlemode}.
The corresponding curves lie in the evanescent sector (i.e. below the light cone) and they are
both bounded by the pair-creation threshold frequency to which they tend for $k\to \infty$.
Further, the coupled modes obey the relation $\Omega_{-}(K)< \Omega_{0}(K)<\Omega_{+}(K)$.
At small wave vectors, however, the two coupled modes behave in a very different way. The
symmetric polariton frequency goes to zero for $k\to 0$ in a way similar to the single layer
mode, although we always have $\Omega_{-}(K)< \Omega_{0}(K)$. Conversely, the $\Omega_{+}(K,\lambda)$
mode stops at
\begin{equation}
    K\equiv K_{\mathrm{lc}}=\frac{\psi^{\rm iv}\left[\frac{1}{\alpha\lambda}\right]}{\sqrt{1-v^2}},
\end{equation}
where, using the parametric expressions in Eq. \eqref{eq:bimodes}, one can also show that the
dispersion relation of the antisymmetric mode becomes tangent to the light cone (i.e. at this
point, the group velocity is $c$).
In the case of two metallic plates
\cite{intravaia05,intravaia07},
the curve corresponding to $\Omega_{+}(K)$ continues above the light cone, indicating a change
of the polaritonic field in the transverse direction from evanescent to propagating.
For the graphene layers considered here, we find no propagating branch for the antisymmetric mode.
This behavior is related to the mathematical properties of $\psi(p)$ in the propagating sector,
below the pair-creation threshold -- a solution would correspond to values for which $\mu$ is
a purely imaginary number while $p<1$.
This feature is similar to what was already observed for the polaritonic modes in a
magneto-dielectric cavity
\cite{Haakh13}.
As in this case, the antisymmetric mode $\Omega_{+}(K)$ can be seen as departing from a continuum
of TE-polarized waves occurring for $ck<\omega$ and are associated with the branch cut in the
reflection coefficient due to the square root of $\kappa$.
Note that the entire light cone ($\mu = 0$) is a trivial, distance-independent solution of
\eqref{eq:resoBi} that describes the antisymmetric polariton.
Due to its properties and in order to preserve the number of modes as a function of the wave
vector, analogous to Ref.~
\cite{Haakh13},
we continue the $\Omega_{+}(K)$ dispersion relation along the light cone from $K=K_{\mathrm{lc}}$
down to zero (see Fig.~\ref{fig:modes}).
Starting from the above considerations, we calculate, in the next section, the contribution
of surface polariton modes to the overall Casimir energy.

\section{Polaritonic contribution to the Casimir energy}
\label{sec:contribution}

In analogy to Eq.~\eqref{eq:casSum}, we define the polaritonic contribution starting from
the zero-point energy associated with the different modes:
\begin{equation}
    E_{\mathrm{pol}}= \sum_{\textbf{k}} \left[ \frac{\hbar\omega_{+}(\mathbf{k},L)}{2} + \frac{\hbar\omega_{-}(\mathbf{k},L)}{2} \right]_{L \rightarrow \infty}^{L}.
\end{equation}
Using our dimensionless variables, this expression can be written as
\begin{equation}
    \frac{E_{\mathrm{pol}}(\lambda)}{E_{\mathrm{N}} }= \int_0^{\infty} K \mathrm{d}K [\Omega_{+}(K,\lambda) + \Omega_{-}(K,\lambda) - 2\Omega_{0}(K)],
\end{equation}
where $E_{\mathrm{N}} = \hbar c k_{\Delta}^3 A/(4 \pi)$. Here, we have already considered
that in the limit $L \rightarrow \infty$ the coupled modes tend to the single layer polariton.
Owing to the implicit nature of the dispersion relations, this  expression does not lend
itself to a simple, straightforward evaluation.
For our analytical and numerical investigations it is convenient to change the integration
variable to $\mu$, which was used as parameter in Eqs. \eqref{eq:singlemode} and \eqref{eq:bimodes}.
Due to the Jacobian $K\mathrm{d}K = \mu \mathrm{d} \mu + \Omega \mathrm{d} \Omega$, the
polaritonic energy can be written as
\begin{multline}
      \frac{E_{\mathrm{pol}}(\lambda)}{E_{\mathrm{N}}}= \frac{1}{3} \left[ \Omega_{+}^3 + \Omega_{-}^3 - 2 \Omega_{0}^3 \right]_{K \rightarrow 0, K_{\mathrm{lc}}}^{K \rightarrow \infty}+ \frac{1}{3} K_{\mathrm{lc}}^3
    \\ + \int_{0}^{\infty} \left[ \Omega_{+}(\mu,\lambda) + \Omega_{-}(\mu,\lambda) - 2 \Omega_{0}(\mu) \right] \mu \mathrm{d} \mu .
\end{multline}
In the first line, the upper limits cancel each other and the lower limit of $\Omega_{0}$
and $\Omega_{-}$ is zero. The remaining lower limit of $\Omega_+$ cancels the second term
leaving us with
\begin{equation}
    \frac{E_{\mathrm{pol}}(\lambda)}{E_{\mathrm{N}}} = \int_{0}^{\infty} \left[ \Omega_{+}(\mu,\lambda) + \Omega_{-}(\mu,\lambda) - 2 \Omega_{0}(\mu) \right] \mu \mathrm{d} \mu.
    \label{eq:intParam}
\end{equation}
This integral allows for a simpler analytical treatment and a robust numerical evaluation.

\begin{figure}
    \centering
    \resizebox{\linewidth}{!}{\includegraphics{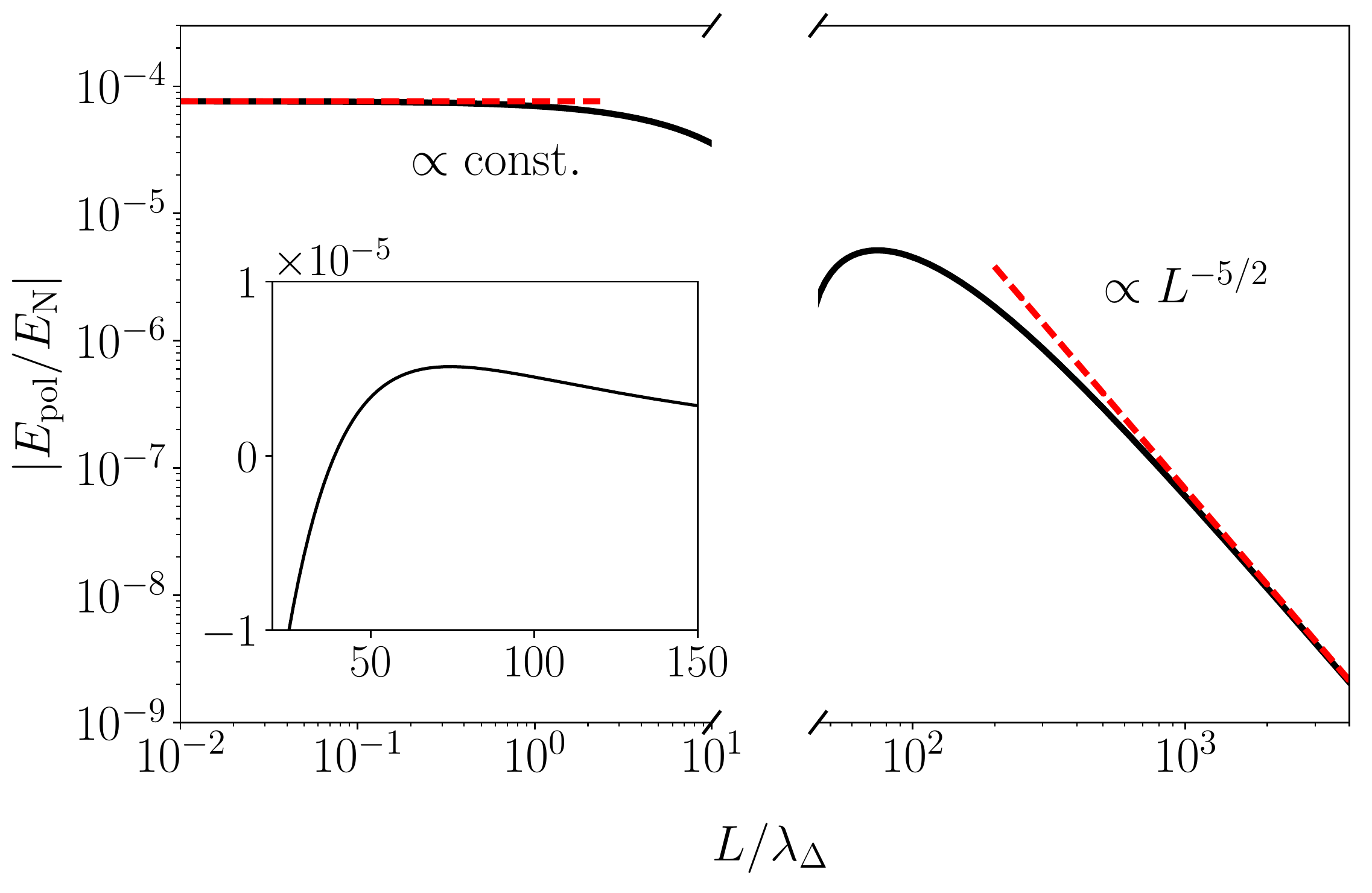}}
    \caption{Absolute value of the polaritonic energy scaled to $E_{\mathrm{N}}$ as a function
        of the separation length in units of $\lambda_{\Delta}$.
        The asymptotes shown as red dashed lines illustrate a scaling of $\propto L^{-5/2}$
						 for large separations and a scaling of $\propto L^{0}$, i.e. a constant for vanishing
						 separations.
             \textit{Inset}: The polaritonic energy exhibits a maximum at $L \approx 74\lambda_{\Delta}$,
						        where the behavior of the polaritonic modes changes from an attracting
										contribution to a repulsing one.
                                    \label{fig:ePl}}
\end{figure}

In Fig.~\ref{fig:ePl}, we depict the polaritonic energy in Eq.~\eqref{eq:intParam} and this
highlights two important features of this contribution to the Casimir energy.
First, in the case of two graphene layers and for $\lambda=L/\lambda_{\Delta}\ll 1$ the energy
tends to a finite negative constant (see also Sec. \ref{DistanceAsymptotes} below). When compared
to the total energy discussed in Sec. \ref{sec:casEnergy} (see Fig. \ref{fig:eCas}), this means
that, contrary to the case of ordinary metals described via a spatially local dielectric model,
where the plasmonic energy negatively diverges and dominates the Casimir interaction at short
separations, for two graphene layers in close proximity, the surface polaritons provide only
a sub-leading contribution to the total energy in Eq. \eqref{eq:lifshitz}.
At this point we would like to emphasize that, different from the total energy, the polaritonic
contribution is less sensitive to the minimal distance constraint discussed in Sec. \ref{sec:casEnergy}
that has been derived from the range of validity of the Dirac model.
Indeed, due to the low value of $v_{\rm F}/c$, the corresponding polaritonic energies, in the
region where they contribute to the Casimir energy ($k\lesssim 1/L$), are limited by the
pair-creation threshold.
The means that the results are reliable for $L\gtrsim (v_{\rm F}/c) \lambda_{\rm Min}$.

Similar to the metallic case, the function $E_{\mathrm{pol}}(\lambda)$ increases with distance between
the graphene layers and reaches a maximum at $L\approx 74\lambda_{\Delta}$.
This indicates that the polaritonic contribution to the total force is attractive for distances
shorter than this value and repulsive at larger separations.
Interestingly, in this latter limit $E_{\mathrm{pol}} \propto L^{-5/2}$, which is the same
scaling that have been observed for surface plasmon polaritons in the metallic case
\cite{intravaia07}.
These findings are also confirmed by the detailed asymptotic analysis reported below.

\subsection{Asymptotic behaviors short and large separations}
\label{DistanceAsymptotes}

From a more mathematical point of view, the difference between the coupled and the isolated
polaritons is due to the behavior of the function $f_{\pm}(x)$.
Since $\psi^{\rm iv}\left[x\gg1\right]\sim 1$, for $\lambda\ll 1$ we obtain the main contribution
to the integral in Eq.~\eqref{eq:intParam} for $\mu\sim \alpha$.
As a result, in the limit $\lambda\to 0$, we have
\begin{equation}
    f_+(\mu \lambda) \approx \frac{1}{\mu \lambda}+\frac{1}{2}\gg 1 ,\quad
    f_-(\mu \lambda) \approx \frac{1}{2}.
\end{equation}
In this limit, the expressions in the integrand of Eq. \eqref{eq:intParam} thus become
distance independent.
Setting $x=\mu/\alpha$, the polaritonic energy approaches a constant given by
\begin{multline}
    \frac{E_{\mathrm{pol}}(\lambda)}{E_{\mathrm{N}} } \xrightarrow{\lambda\to 0} \alpha^{2}\int_0^{\infty}  \mathrm{d}x \,x\left[ \sqrt{\frac{(\alpha v)^2 x^2 + 1}{1-v^2}} \right.
    \\ \left. +  \sqrt{\frac{(\alpha v)^2 x^2 + \psi^{\rm iv}\left[x/2\right]^2}{1-v^2}}- 2  \sqrt{\frac{(\alpha v)^2 x^2 + \psi^{\rm iv}\left[x\right]^2}{1-v^2}} \right].
\end{multline}
This expression scales with $\alpha^{2}$ and is only weakly depending on $v$.
Using $\alpha = 1/137$ and $v = 1/300$ gives $E_{\mathrm{pol}}/ E_{\mathrm{N}} \approx -7.6\times 10^{-5}$.

For larger separations, the dispersion relations of the coupled polaritonic modes are
very close to the uncoupled one.
The difference is controlled by the small parameter $e^{-\mu\lambda}$ which is significant
only for $\mu \lesssim 1/\lambda$. In the limit $1/\lambda \to 0$ we may then neglect the
first term under the square roots in Eqs.~\eqref{eq:singlemode} and \eqref{eq:bimodes} and
consider that $\psi^{\rm iv}(x \rightarrow 0) \approx \sqrt{3x}/2$.
We then have
\begin{equation}
    \Omega_{\pm}(\mu,\lambda) - \Omega_{0}(\mu) \approx \sqrt{\frac{3\frac{\mu}{\alpha}}{1-v^2}}\frac{\sqrt{f_{\pm}(\mu \lambda)} - \sqrt{f_{0}(\mu \lambda)}}{2}.
\end{equation}
Inserting the previous expressions in Eq.~\eqref{eq:intParam} and employing a change of
variable $x = \mu \lambda$, it is straightforward to show that, for large distances, the
polaritonic energy goes as
\begin{equation}
    \frac{E_{\mathrm{pol}}(\lambda\gg 1)}{E_{\mathrm{N}} } \approx \sqrt{\frac{3}{4 \alpha (1-v^2)}} \frac{C}{\lambda^{5/2}},
\end{equation}
where $C$ is a numerical constant given by
\begin{align}
    C &= \int_0^{\infty} \mathrm{d} x\, x^{\frac{3}{2}} \Bigg\{ \sqrt{\frac{1+ \tanh \left[\frac{x}{2} \right]}{2}} + \sqrt{\frac{1+ \coth \left[\frac{x}{2} \right]}{2}} - 2 \Bigg\} \nonumber\\
    &\approx 0.2132.
\end{align}
The polaritonic modes thus give a contribution to the energy which is positive and vanishes
slower than the total energy ($E(L) \propto L^{-5}$).

\section{Discussion and conclusions}
\label{sec:conclusion}

In this work, we have analyzed in detail the contribution of surface polaritons to the
zero-temperature Casimir interaction between two parallel graphene layers that are separated
by vacuum and are described by the Dirac model. In our description we include a small gap in
the band structure of graphene that accounts for the effect of strain or other experimental
conditions which can lead to a breaking of symmetry in the material's lattice structure.

Specifically, we have derived parametric expressions for the two coupled surface modes that
result from the hybridization of the two single-layer polaritons and have calculated their
contribution to the Casimir energy.
For the parameters considered ($T=0$ and vanishing chemical potential), the system only allows for TE-polarized
surface resonances. Despite the complexity of the expressions, our approach has allowed for
a detailed analytical description of the polaritonic energy.
We have shown that all modes are associated with an evanescent field: the dispersion relations
of the single-layer $\omega_{0}(k)$ and the symmetric coupled mode $\omega_{-}(k)$ tend to
zero for $k\to 0$; the remaining antisymmetric coupled mode $\omega_{+}(k)$ becomes instead
tangent to the light cone for a positive and distance dependent value of the wave vector.
A similar behavior was observed in the case of the surface polaritons occurring in a
magneto-dielectric cavity
\cite{Haakh13}.

Further, we have analyzed the behavior of the polaritonic contribution for small and large
separations $L$ between the layers and have contrasted the resulting expressions with those
for the total Casimir energy.
Contrary to ordinary metals, for which the polaritonic contribution describes the nonretarded
short-distance behavior (van der Waals limit), the total energy for graphene scales $\propto L^{-3}$
while the surface modes' energy tends to a constant.
Due to the band gap, at large distances the total zero-temperature Casimir energy exhibits
an unusual $L^{-5}$ behavior, while the polaritonic energy scales as $\propto L^{-5/2}$,
similar to that found for the corresponding contribution in the case of two metallic
plates
\cite{intravaia05,bordag06,intravaia07}.
In analogy with this last configuration, the polaritonic Casimir energy has the interesting
property to exhibit a maximum at a distance  which for graphene scales as the inverse of the
bandgap energy.
Consequently, the polaritonic force changes sign and being attractive at short separations
and becoming repulsive at large separations. Nonetheless, the total Casimir force remains
attractive throughout.

Our results show that, in the technologically interesting limit of very small separations
(van der Waals solids), graphene's surface resonances can behave in a quite unusual way
relative to the analogous situation for ordinary metals.

\section{Acknowledgments}
\label{sec:acknowledgments}

We are indebted with I.~G.~Pirozhenko, D.~Reiche and M.~Oelschl\"ager for insightful and fruitful discussions.
Further, we acknowledge support by the Deutsche Forschungsgemeinschaft (DFG)
through SFB 951 HIOS (B10, Project No. 182087777).
FI acknowledges financial support from the DFG through the DIP program (Grant No. SCHM 1049/7-1).

\appendix

\section{The impact of graphene's onto-electronic properties on the Casimir energy}
\label{sec:casEnergCalc}

As explained in the main text, in order to investigate the Casimir energy between
two parallel graphene sheets, it is convenient to use the function $\eta(\alpha,v,\lambda)$,
which compares the Casimir energy with the energy between two perfectly reflecting
surfaces.
With the change of variable $\tilde{\xi}= 2 \lambda \xi \lambda_{\Delta}/c$ and
$\tilde{k} = 2 \lambda k\lambda_{\Delta}$, we may write the reflection coefficients
as
\begin{subequations}
    \begin{gather}
        r_{\textrm{TE}}(\imath\tilde{\xi}, \tilde{k}) = - \frac{2\lambda\alpha \varphi\left(\frac{\tilde{\rho}}{2\lambda}\right)}{\tilde{\kappa} + 2\lambda\alpha \varphi\left(\frac{\tilde{\rho}}{2\lambda}\right)},
        \\
        r_{\textrm{TM}}(\imath\tilde{\xi}, \tilde{k}) = \frac{2\lambda\alpha\tilde{\kappa} \varphi\left(\frac{\tilde{\rho}}{2\lambda}\right)}{2\lambda\alpha\tilde{\kappa} \varphi\left(\frac{\tilde{\rho}}{2\lambda}\right) + \tilde{\rho}^2},
    \end{gather}
\end{subequations}
where $\tilde{\kappa}=\sqrt{\tilde{k}^2 + \tilde{\xi}^2}$, $\tilde{\rho} = \sqrt{\tilde{\xi}^2 + v^2 \tilde{k}^2} $ and
\begin{equation}
    \varphi(x)\equiv-\psi(\imath x) =\left[ 1 + \left(x - \frac{1}{x} \right)\arctan\left(x \right) \right].
\end{equation}
In this case, the function $\eta(\alpha,v,\lambda)$ can be written as follows
\begin{multline}
    \eta(\alpha,v,\lambda) = - \frac{45}{2 \pi^{4}}\int_0^{\infty} \mathrm{d}\tilde{\xi} \int_0^{\infty} \mathrm{d}\tilde{k}\, \tilde{k} \\
    \sum_{\sigma} \ln \left[ 1 -  r^2_{\mathrm{\sigma}}(\imath \tilde{\xi}, \tilde{k})_{\vert\lambda \rightarrow 0} e^{-\tilde{\kappa}} \right]
    \label{eq:constShort}
\end{multline}
Due to the exponential in the integrand the dominant contributions arise for $1\gtrsim\tilde{\kappa}>\tilde{\rho}$.

In the limit $\lambda\to 0$ we can, therefore, consider the limit $\varphi(x) \approx x \pi/2$
that is obtained for $x\to \infty$.
In this case, the resulting expressions for the reflection coefficients are the
same as those that are obtained for the limit $\Delta \to 0$
\begin{gather}
    r_{\textrm{TE}}(\imath\tilde{\xi}, \tilde{k}) \approx - \frac{\alpha \frac{\pi}{2}\tilde{\rho}}{\tilde{\kappa} + \alpha \frac{\pi}{2}\tilde{\rho}},
    \quad
    r_{\textrm{TM}}(\imath\tilde{\xi}, \tilde{k})\approx \frac{\alpha\tilde{\kappa}\frac{\pi}{2}}{\alpha\tilde{\kappa} \frac{\pi}{2} + \tilde{\rho}} .
\end{gather}
As a consequence, the function $ \eta(\alpha,v,\lambda)$ does not depend on $\lambda$.
The above expressions also show that, in this limit, the TM contribution is larger
than the TE contribution.
In order to obtain an analytically tractable expression, we introduce polar coordinates
in the $\tilde{\xi}$-$\tilde{k}$-plane, $\tilde{\xi}=h \sin[\phi]$ and $\tilde{k}=h \cos[\phi]$,
and simplify the integration over the angle through the change of variable $x = \sin\phi$.
The integral can then be solved analytically but features rather lengthy expressions,
the full form of which we do not want to give here.
We have
\begin{align}
    \eta(\alpha,v,0)\approx \frac{45}{\pi^{4}}\left[g_{\mathrm{TE}}(\alpha,v) + g_{\mathrm{TM}}(\alpha,v) \right],
\end{align}
where
\begin{align}
    g_{\mathrm{TE}}(\alpha,v) = \int_0^{1} \mathrm{d}x \left( \frac{\frac{\alpha \pi}{2}\sqrt{x^2(1-v^2) + v^2}}{1 + \frac{\alpha \pi}{2}\sqrt{x^2(1-v^2) + v^2}} \right)^2
\end{align}
and
\begin{align}
    g_{\mathrm{TM}}(\alpha,v) = \int_0^{1} \mathrm{d}x \left( \frac{1}{1 + \frac{2}{\alpha \pi}\sqrt{x^2(1-v^2) + v^2}} \right)^2.
\end{align}
Further, we observe that for very small $\alpha$, we can Taylor-expand the expression
\begin{align}
    \eta(\alpha, v, 0) \stackrel{\alpha \ll 1}{\approx} \frac{45}{4 \pi^2}\alpha ^2 \left(\frac{1}{3} \left(2 v^2+1\right)+\frac{\arctan\left(\sqrt{\frac{1 -v^2}{v^2}}\right)}{\sqrt{v^2 \left(1-v^2\right)}}\right)
\end{align}
and obtain a scaling $\eta \propto \alpha^2$.

For $\lambda\gg 1$, we can use the approximation $\varphi(x) \approx 4 x^2/3$ valid for $x\ll1$.
The reflection coefficients then become
\begin{gather}
    r_{\textrm{TE}}(\imath\tilde{\xi}, \tilde{k}) \approx - \frac{2}{3}\frac{\alpha}{\lambda}\frac{\tilde{\rho}^{2}}{\tilde{\kappa}},
    \quad
    r_{\textrm{TM}}(\imath\tilde{\xi}, \tilde{k}) \approx \frac{2}{3}\frac{\alpha}{\lambda}\tilde{\kappa}
\end{gather}
Similarly to the previous case, we consequently have
\begin{align}
    \eta(\alpha,v,\lambda) &\stackrel{\lambda\gg 1}{\approx}  \frac{\alpha^{2}}{\lambda^{2}}\frac{10}{\pi^{4}}  \int_0^{\infty} \mathrm{d}\tilde{\xi} \int_0^{\infty} \mathrm{d}\tilde{k}\, \tilde{k}
    \left[\tilde{\kappa}^{2}-\frac{\tilde{\rho}^{4}}{\tilde{\kappa}^{2}}\right] e^{-\tilde{\kappa}}\nonumber\\
    &= \frac{240 \alpha^2}{\pi^4 \lambda^{2}}\left[ 1 + \frac{1}{15}\left( 3 + 4v^2 + 3v^4 \right) \right].
    \label{eq:constlong}
\end{align}


\begin{thebibliography}{10}

\bibitem{casimir48}
H.~B.~G. Casimir, On the attraction between two perfectly conducting plates,
  Proc. K. Ned. Akad. Wet. {\bf 51},  793  (1948).

\bibitem{van-kampen68}
N. van Kampen, B. Nijboer, and K. Schram, On the macroscopic theory of Van Der
  Waals forces, Phys. Lett. A {\bf 26},  307  (1968).

\bibitem{henkel04}
C. Henkel, K. Joulain, J.-P. Mulet, and J.-J. Greffet, Coupled surface
  polaritons and the Casimir force, Phys. Rev. A {\bf 69},  023808  (2004).

\bibitem{intravaia05}
F. Intravaia and A. Lambrecht, Surface Plasmon Modes and the Casimir Energy,
  Phys. Rev. Lett. {\bf 94},  110404  (2005).

\bibitem{barton05}
G. Barton, Casimir effects for a flat plasma sheet: I. Energies, J. Phys. A
  Math. Gen. {\bf 38},  2997  (2005).

\bibitem{bordag06}
M. Bordag, The Casimir effect for thin plasma sheets and the role of the
  surface plasmons, J. Phys. A Math. Gen. {\bf 39},  6173  (2006).

\bibitem{intravaia07}
F. Intravaia, C. Henkel, and A. Lambrecht, Role of surface plasmons in the
  Casimir effect, Phys. Rev. A {\bf 76},  033820  (2007).

\bibitem{haakh16}
H.~R. Haakh, S. Faez, and V. Sandoghdar, Polaritonic normal-mode splitting and
  light localization in a one-dimensional nanoguide, Phys. Rev. A {\bf 94},
  053840  (2016).

\bibitem{economou69}
E.~N. Economou, Surface Plasmons in Thin Films, Phys. Rev. {\bf 182},  539
  (1969).

\bibitem{joulain05}
K. Joulain, J.-P. Mulet, F. Marquier, R. Carminati, and J.-J. Greffet, Surface
  electromagnetic waves thermally excited: Radiative heat transfer, coherence
  properties and Casimir forces revisited in the near field, Surface Science
  Reports {\bf 57},  59   (2005).

\bibitem{pitarke07}
J.~M. Pitarke, V.~M. Silkin, E.~V. Chulkov, and P.~M. Echenique, Theory of
  surface plasmons and surface-plasmon polaritons, Rep. Prog. Phys. {\bf 70},
  1  (2007).

\bibitem{maier07}
S.~A. Maier, {\em Plasmonics: fundamentals and applications} (Springer Science
  and Business Media, New York, 2007).

\bibitem{nie97}
S. Nie and S. Emory, Probing single molecules and single nanoparticles by
  surface-enhanced Raman scattering, Science {\bf 275},  1102  (1997).

\bibitem{ebbesen98}
T. Ebbesen, H. Lezec, H. Ghaemi, T. Thio, and P. Wolff, Extraordinary optical
  transmission through sub-wavelength hole arrays, Nature {\bf 391},  667
  (1998).

\bibitem{torma15}
P. T{\"o}rm{\"a} and W.~L. Barnes, Strong coupling between surface plasmon
  polaritons and emitters: a review, Rep. Prog. Phys. {\bf 78},  013901
  (2015).

\bibitem{Intravaia12a}
F. Intravaia, P.~S. Davids, R.~S. Decca, V.~A. Aksyuk, D. L\'opez, and D.~A.~R.
  Dalvit, Quasianalytical modal approach for computing Casimir interactions in
  periodic nanostructures, Phys. Rev. A {\bf 86},  042101  (2012).

\bibitem{intravaia13}
F. Intravaia {\it et~al.}, {Strong Casimir force reduction through metallic
  surface nanostructuring}, Nat. Commun. {\bf 4},  2515  (2013).

\bibitem{fialkovsky11}
I.~V. Fialkovsky, V.~N. Marachevsky, and D.~V. Vassilevich, Finite-temperature
  Casimir effect for graphene, Phys. Rev. B {\bf 84},  035446  (2011).

\bibitem{banishev13}
A.~A. Banishev, G.~L. Klimchitskaya, V.~M. Mostepanenko, and U. Mohideen,
  Demonstration of the Casimir Force between Ferromagnetic Surfaces of a
  Ni-Coated Sphere and a Ni-Coated Plate, Phys. Rev. Lett. {\bf 110},  137401
  (2013).

\bibitem{Biehs14}
S.-A. Biehs and G.~S. Agarwal, Anisotropy enhancement of the Casimir-Polder
  force between a nanoparticle and graphene, Phys. Rev. A {\bf 90},  042510
  (2014).

\bibitem{klimchitskaya15}
G.~L. Klimchitskaya and V.~M. Mostepanenko, Comparison of hydrodynamic model of
  graphene with recent experiment on measuring the Casimir interaction, Phys.
  Rev. B {\bf 91},  045412  (2015).

\bibitem{klimchitskaya15a}
G.~L. Klimchitskaya and V.~M. Mostepanenko, Origin of large thermal effect in
  the Casimir interaction between two graphene sheets, Phys. Rev. B {\bf 91},
  174501  (2015).

\bibitem{bordag16}
M. Bordag, I. Fialkovskiy, and D. Vassilevich, Enhanced Casimir effect for
  doped graphene, Phys. Rev. B {\bf 93},  075414  (2016).

\bibitem{Henkel18}
C. Henkel, G.~L. Klimchitskaya, and V.~M. Mostepanenko, Influence of the
  chemical potential on the Casimir-Polder interaction between an atom and
  gapped graphene or a graphene-coated substrate, Phys. Rev. A {\bf 97},
  032504  (2018).

\bibitem{Klimchitskaya18a}
G.~L. Klimchitskaya and V.~M. Mostepanenko, Graphene may help to solve the
  Casimir conundrum in indium tin oxide systems, Phys. Rev. B {\bf 98},  035307
   (2018).

\bibitem{Geim13}
A.~K. Geim and I.~V. Grigorieva, Van der Waals heterostructures, Nature {\bf
  499},  419 EP   (2013).

\bibitem{Gobre13}
V.~V. Gobre and A. Tkatchenko, Scaling laws for van der Waals interactions in
  nanostructured materials, Nat. Commun. {\bf 4},    (2013).

\bibitem{Novoselov16}
K.~S. Novoselov, A. Mishchenko, A. Carvalho, and A.~H. Castro~Neto, 2D
  materials and van der Waals heterostructures, Science {\bf 353},    (2016).

\bibitem{Ambrosetti16}
A. Ambrosetti, N. Ferri, R.~A. DiStasio, and A. Tkatchenko, Wavelike charge
  density fluctuations and van der Waals interactions at the nanoscale, Science
  {\bf 351},  1171  (2016).

\bibitem{Woods16}
L.~M. Woods, D.~A.~R. Dalvit, A. Tkatchenko, P. Rodriguez-Lopez, A.~W.
  Rodriguez, and R. Podgornik, Materials perspective on Casimir and van der
  Waals interactions, Rev. Mod. Phys. {\bf 88},  045003  (2016).

\bibitem{mccann12}
E. McCann,  in {\em Graphene Nanoelectronics}, {\em NanoScience and
  Technology}, edited by H. Raza (Springer, Berlin Heidelberg, 2012), pp.\
  237--275.

\bibitem{Werra16}
J.~F.~M. Werra, F. Intravaia, and K. Busch, TE resonances in
  graphene-dielectric structures, J. Opt. {\bf 18},  034001  (2016).

\bibitem{Zhou07}
S.~Y. Zhou {\it et~al.}, Substrate-induced bandgap opening in epitaxial
  graphene, Nat. Mater. {\bf 6},  770  (2007).

\bibitem{Giovannetti07}
G. Giovannetti, P.~A. Khomyakov, G. Brocks, P.~J. Kelly, and J. van~den Brink,
  Substrate-induced band gap in graphene on hexagonal boron nitride: \textit{Ab
  initio} density functional calculations, Phys. Rev. B {\bf 76},  073103
  (2007).

\bibitem{Chen14}
Z.-G. Chen {\it et~al.}, Observation of an intrinsic bandgap and Landau level
  renormalization in graphene/boron-nitride heterostructures, Nat. Commun. {\bf
  5},    (2014).

\bibitem{Jung15}
J. Jung, A.~M. DaSilva, A.~H. MacDonald, and S. Adam, Origin of band gaps in
  graphene on hexagonal boron nitride, Nat. Commun. {\bf 6},    (2015).

\bibitem{chaichian12}
M. Chaichian, G.~L. Klimchitskaya, V.~M. Mostepanenko, and A. Tureanu, Thermal
  Casimir-Polder interaction of different atoms with graphene, Phys. Rev. A
  {\bf 86},  012515  (2012).

\bibitem{Falkovsky07}
L.~A. Falkovsky and A.~A. Varlamov, Space-time dispersion of graphene
  conductivity, Eur. Phys. J. B {\bf 56},  281  (2007).

\bibitem{Falkovsky08}
L.~A. Falkovsky, Optical properties of graphene, J. Phys. Conf. Ser. {\bf 129},
   012004  (2008).

\bibitem{bordag15}
M. Bordag and I.~G. Pirozhenko, Surface plasmons for doped graphene, Phys. Rev.
  D {\bf 91},  085038  (2015).

\bibitem{lifshitz56}
E. Lifshitz, The Theory of Molecular Attractive Force between Solids, Soviet
  Phys. JETP {\bf 2},  73  (1956).

\bibitem{genet03}
C. Genet, A. Lambrecht, and S. Reynaud, Casimir force and the quantum theory of
  lossy optical cavities, Phys. Rev. A {\bf 67},  043811  (2003).

\bibitem{klimchitskaya13}
G.~L. Klimchitskaya and V.~M. Mostepanenko, van der Waals and Casimir
  interactions between two graphene sheets, Phys. Rev. B {\bf 87},  075439
  (2013).

\bibitem{bordag14}
M. Bordag and I.~G. Pirozhenko, Transverse-electric surface plasmon for
  graphene in the Dirac equation model, Phys. Rev. B {\bf 89},  035421  (2014).

\bibitem{mikhailov07}
S.~A. Mikhailov and K. Ziegler, New Electromagnetic Mode in Graphene, Phys.
  Rev. Lett. {\bf 99},  016803  (2007).

\bibitem{Haakh13}
H.~R. Haakh and F. Intravaia, Mode structure and polaritonic contributions to
  the Casimir effect in a magnetodielectric cavity, Phys. Rev. A {\bf 88},
  052503  (2013).

\end{thebibliography}
\end{document}